\documentclass[12pt]{article}

\newcommand{\be}{\begin{equation}}
\newcommand{\ee}{\end{equation}}

\begin{document}
\begin{center}
\large{\textbf{\textbf{On the Stability of Coherent States for Pais-Uhlenbeck
Oscillator}}}\\
\end{center}
\begin{center}
Souvik Pramanik and Subir Ghosh \\
Physics and Applied Mathematics Unit, Indian Statistical
Institute\\
203 B. T. Road, Kolkata 700108, India \\
\end{center}\vspace{0.5cm}

\begin{center}
{\textbf{Abstract}}
\end{center}
We have constructed coherent states for the higher derivative Pais-Uhlenbeck
Oscillator. In the process
we have suggested a novel way to construct coherent states for the oscillator
having only negative
energy levels. These coherent states have negative energies in general but their
coordinate and momentum expectation values
and dispersions behave in an identical manner as that of normal (positive
energy) oscillator. The coherent states
for the Pais-Uhlenbeck Oscillator have constant dispersions and a modified
Heisenberg Uncertainty Relation. Moreover,
under reasonable assumptions on parameters these coherent states can have
positive energies.

\vskip .5cm

{\it{Introduction}}: The Pais-Uhlenbeck Oscillator (PUO) \cite{pu} can act as a
prototype toy model of higher derivative (covariant) theories  of gravity
\cite{hd} since both are plagued by ghost excitations. In PUO the ghost problem
is manifested in two complimentary ways:
 (i) The Hamiltonian can be bounded below but the system will contain negative
norm states; (ii) the system will consist only of positive norm states but
the Hamiltonian is not bounded below.
 In \cite{ben,ben2} the authors solve both the problems by treating PUO as a
non-Hermitian
$PT$-symmetric quantum system. However some questions have been raised in
\cite{sm,sm2,mos} and in \cite{mos} an
alternative consistent formulation of PUO has also been suggested. We follow the
spirit of Smilga \cite{sm,sm2} who
suggests that the ghost induced non-unitarity is not a serious problem in the
context of PUO since the theory is free. Indeed an important question is what
will happen in presence
of interactions. We will comment on this at the end. However we point out that
 interactions will lead to an  extended version of PUO. We are not
dealing with this non-minimal form of PUO at present.
In the present paper we follow the  option (ii) stated above. We concentrate on
the
stability problem arising from the presence
of negative energy levels. Specifically we show that coherent states,
constructed in the
Generalized Coherent State (GCS) framework, pioneered by Klauder, Gazeau and
others \cite{gcs,klauder,gcs1}, are stable with
positive energy under reasonable assumptions on the parameters. In the process
we have
suggested a simple but novel way of accommodating systems with non-positive
energy
spectrum in GCS scheme. Recently we have applied GCS scheme \cite{sp} in
extended Harmonic Oscillator models satisfying Generalized Uncertainty Principle
and in non-linear oscillator model \cite{ghosh}.

{\it{Pais-Uhlenbeck Oscillator in Hamiltonian framework}}: The dynamics of
classical PUO is governed by the following equation of motion:
\be
z^{(4)}+(\Omega^2+\omega^2 )z^{(2)}+\Omega^2\omega^2 z=0,
\label{z}
\ee
where $z^{(k)}$ denotes $k$-th time derivative of the   real dynamical variable
$z$ and $\Omega,\omega$ and positive real parameters. In the present work we
will concentrate on the non-degenerate case $\Omega>\omega $ where
$e^{\pm i\Omega t}$ and $e^{\pm i\omega t}$ are the solutions of (\ref{z}). The
dynamics can be obtained from the higher derivative Lagrangian,
\be
L=\frac{1}{2}[\ddot z^2-(\Omega^2+\omega^2 )\dot z^2+\omega ^2\Omega^2
z^2],
\label{xz}
\ee
or its classically equivalent counterpart,
\be
L=\frac{1}{2}[\dot q^2-(\Omega^2+\omega^2 ) q^2+\omega ^2\Omega^2
z^2]+\lambda (\dot z-q).
\label{xz1}
\ee
The Lagrangian (\ref{xz1}) is quadratic at the expense of an additional real
variable $q$ besides $z$. This alternative is suitable for us. Let us follow
Dirac's Hamiltonian formalism for constraint systems \cite{dir}. For the PUO
this is discussed in \cite{mann}.

The conjugate momenta $p_q=(\partial L)/(\partial \dot q )=\dot q
,~p_z=(\partial L)/(\partial \dot z )=\lambda ,~p_\lambda =(\partial
L)/(\partial \dot \lambda )=0$ show that there are two
Second Class Constraints (non-commuting in the Poisson Bracket sense)
$\psi_1=p_z-\lambda \approx 0,~\psi_2=p_\lambda \approx 0$ with
$\{\psi_1,\psi_2\}=-1$. All the variables $q,z,\lambda$
are considered to be canonical and independent. One can impose the constraints
strongly in further analysis provided Dirac Brackets are used in place of
Poisson Brackets. However, in the present system
it is trivial to see that imposition of the constraints will reduce the set of
variables to $q,p_q,z,p_z$ and their Dirac Brackets will be identical to their
original canonical Poisson Brackets. That is $q,z$
continue to be independent and canonical.

The canonical
Hamiltonian, with the constraints imposed, is,
\be
H=\dot qp_q+\dot zp_z+\dot \lambda p_\lambda -L
$$$$=\frac{1}{2}[p_q^2+2qp_z+(\Omega^2+\omega^2 )q^2-\omega ^2\Omega^2
z^2].
\label{hxz}
\ee
The effect of the constraints is seen in the crossterm $qp_z$ in $H$ (4) and
still later in (32) where we explicitly demonstrate that the Coherent State
constructed here does indeed satisfy
the constraints.

{\it{Canonical transformation to free system of two oscillators}}:  $H$ is
decoupled by exploiting the linear canonical transformation,
\be
X=\frac{p_z+\Omega^2q}{\Omega{\sqrt{\Omega^2-\omega^2}}},~x=\frac{
p_q+\Omega^2z}{{\sqrt{\Omega^2-\omega^2}}},$$$$
P=\frac{\Omega(p_q+\omega^2z)}{{\sqrt{\Omega^2-\omega^2}}},
~p=\frac{p_z+\omega^2q}{{\sqrt{\Omega^2-\omega^2}}},
\label{xpx}
\ee
yielding,
\be
H=\frac{1}{2}(P^2+\Omega^2X^2)-\frac{1}{2}(p^2+\omega^2x^2).
\label{hh1}
\ee
Quantization of the individual oscillators leads to the energy spectrum
$\Omega(n_1+\frac{1}{2})-\omega(n_2+\frac{1}{2})$ that is clearly not positive
definite due to the presence of the $\omega $ oscillator. However both the
oscillators live in
a positive norm Hilbert space. On the other hand one can have an equivalent and
alternative setup where the
$\omega $ oscillator will yield positive energy states but the states will have
negative norm. These two scenarios result from the inherent ghost problem of
higher derivative theories, PUO being an
example. These issues are discussed in detail in \cite{ben,ben2} where it is stressed
that both choices need to be avoided.  However, as commented in \cite{sm,sm2}, the
absence of a lower bound in
energy does not cause a serious problem unless non-linear interactions are
involved.

{\it{Coherent state for negative energy oscillator}}: Our subsequent analysis is
in agreement with conclusions of \cite{sm,sm2}. We will show that Coherent States
for PUO can represent  stable
states with sensible and unambiguous dispersions, even though the  energy may
not be  positive definite. In fact the Coherent State
energy can be positive for the non-degenerate case studied here with some
natural choice of parameters.
 We start by providing a brief description of the GCS for positive energy
Harmonic Oscillator. Quite
obviously generalization of Coherent States is not required here but we retain
the notations of GCS as in \cite{gcs1}. The Harmonic Oscillator Hamiltonian is,
\be
H=\frac{P^2}{2M}+\frac{M\Omega ^2X^2}{2}.
\label{h0}
\ee
The Fock space is defined by
\be
A\mid n>=\sqrt n\mid n-1>~,~~A^\dagger \mid n>=\sqrt {n+1}\mid n+1>.
\label{n1}
\ee
$H$ is diagonalized to
\be
H=\Omega (A^\dagger A+\frac{1}{2}),
\label{HH2}
\ee
by the creation (annihilation) operator $A^\dagger $ ($A$),
\be
A={\sqrt{\frac{M\Omega}{2}}}X+i\frac{P}{{\sqrt{2M\Omega}}},~A^\dagger={\sqrt{
\frac{M\Omega}{2}}}X-i\frac{P}{{\sqrt{2M\Omega}}},~[A,A^\dagger ]=1.
 \label{AA}
\ee
The GCS is defined as \cite{gcs1}
\be
\mid J,\Gamma >=\frac{1}{N(J)}\sum_{n=0}^\infty \frac{J^{(n/2)}e^{-i\Gamma
E_n}}{{\sqrt{R_n}}}\mid  n>,~R_n=E_1E_2...E_n,~E_n=n
\label{J}
\ee
where $J$ is related to energy and $\Gamma \sim \Omega t$ \cite{gcs1}. The
normalization condition yields $\frac{1}{N(J)^2}\sum_{n=0}^\infty
\frac{J^n}{R_n}=1$. Expectation values of the dynamical variables $X,P$ are
computed easily by using the relations,
\be
X=\frac{A+A^\dagger}{{\sqrt{2M\Omega}}},~~P=-i{\sqrt{\frac{M\Omega}{2}}}
(A-A^\dagger).
\label{x1}
\ee
This yields, for example,
\be
<J,\Gamma \mid A\mid J,\Gamma >=\frac{1}{N^2(J)}\sum_{n,m=0}^\infty
\frac{J^{(n+m)/2)}e^{-i\Gamma (E_m-E_n)}}{{\sqrt{R_mR_n}}}<m\mid A\mid n>$$$$
=\sqrt Je^{-i\Gamma },
\label{A1}
\ee
leading to,
\be
<X>={\sqrt{\frac{2J}{M\Omega}}}cos~\Gamma,~~<P>=-{\sqrt{2M\Omega J}}sin~\Gamma .
\label{X2}
\ee
Finally we come to the coherent state construction for negative energy
oscillator. This is a new approach not considered before. Recall
that we will opt for the second alternative scheme (ii) the system will
consist only of positive norm states but
the Hamiltonian is not bounded below. Hence we  suggest that the same
construction can be applied for
the
negative energy Hamiltonian,
\be
h=-(\frac{p^2}{2m}+\frac{m\omega ^2x^2}{2}),
\label{h1}
\ee
with
\be
a={\sqrt{\frac{m\omega}{2}}}x+i\frac{p}{{\sqrt{2m\omega}}},~a^\dagger={\sqrt{
\frac{m\omega}{2}}}x-i\frac{p}{{\sqrt{2m\omega}}},~[a,a^\dagger ]=1,
\label{X21}
\ee
leading to
\be
h=-\omega (a^\dagger a+\frac{1}{2}).
\label{haa}
\ee
Once again the Fock space is
\be
a\mid n>=\sqrt n\mid n-1>~,~~a^\dagger \mid n>=\sqrt {n+1}\mid n+1>.
\label{n2}
\ee
We propose the GCS to be,
\be
\mid j,\gamma >=\frac{1}{N(j)}\sum_{n=0}^\infty \frac{j^{(n/2)}e^{-i\gamma
(-e_n)}}{{\sqrt{\rho_n}}}\mid n>,~~\rho_n=e_1e_2...e_n=(-1)^nR_n,~e_n=-n.
\label{j}
\ee
The degrees of freedom are
\be
x=\frac{a+a^\dagger}{{\sqrt{2m\omega}}},~~p=-i{\sqrt{\frac{m\omega}{2}}}
(a-a^\dagger).
\label{x3}
\ee
Expectation values of $a,a^\dagger$ are different from the normal case
(13),
\be
<j,\gamma \mid a\mid j,\gamma >=\frac{1}{N^2(J)}\sum_{n,m=0}^\infty
\frac{j^{(n+m)/2)}e^{i\Gamma (e_m-e_n)}}{{\sqrt{\rho_m\rho_n}}}<m\mid a\mid
n>$$$$
=-i\sqrt je^{i\gamma }
\label{aa}
\ee
leading to
\be
<x>={\sqrt{\frac{2j}{m\omega}}}sin~\gamma,~~<p>=-{\sqrt{2m\omega j}}~cos~\gamma
.
\label{X1}
\ee
These can be contrasted with (\ref{X2}). {\it{It is interesting to note that the
$90$ degree phase shift
between (\ref{X1}) and (\ref{X2}) is reminiscent of the $90$ degree rotation to
implement the complexification of
one of the coordinates in }} \cite{ben,ben2}.

One finds the dispersions and the
uncertainty relation,
\be
<x^2>=\frac{1}{2m\omega}(1+4j~sin^2~\gamma
),~~<p^2>=\frac{m\omega}{2}(1+4j~cos^2~\gamma ),
\label{x22}
\ee
\be
(\Delta x)^2=<x^2>-<x>^2=\frac{1}{2m\omega },~~(\Delta
p)^2=<p^2>-<p>^2=\frac{m\omega}{2 }$$$$ (\Delta) x^2(\Delta p)^2=1/4.
\label{dx}
\ee
For the normal case similar well known relations are,
\be
<X^2>=\frac{1}{2M\Omega}(1+4J~cos^2~\Gamma
),~~<P^2>=\frac{M\Omega}{2}(1+4J~sin^2~\gamma ),
\label{X23}
\ee
\be
(\Delta X)^2=<X^2>-<X>^2=\frac{1}{2M\Omega },~~(\Delta
P)^2=<P^2>-<P>^2=\frac{M\Omega}{2 },$$$$ (\Delta X)^2 (\Delta P)^2=1/4.
\label{dX}
\ee
The above are some of our major results indicating that as far as stability of
the coherent states are considered,
an oscillator with only negative energy levels behaves similarly as a normal
positive energy oscillator.

Identical equations of motion satisfy the Correspondence Principle,
\be
<\ddot X>=-\Omega ^2<X>,~~<\ddot x>=-\omega ^2<x>
\label{eq}
\ee
Finally energies of the respective GCS for normal and ghost oscillators are,
\be
<H>=\frac{<P^2>}{2M}+\frac{M\Omega
^2<X^2>}{2}=\frac{\Omega}{2}(1+2J),$$$$<h>=-(\frac{<p^2>}{2M}+\frac{m\omega
^2<x^2>}{2})=-\frac{\omega}{2}(1+2j).
\label{hh3}
\ee

{\it{Coherent state for Pais-Uhlenbeck Oscillator}}: Since PUO is a combination
of two
non-interacting positive and negative energy oscillator (\ref{hh1}), it is
natural to consider the coherent state
as a direct product of $\mid J,\Gamma >$ and $\mid j,\gamma >$, coherent states
of the positive and negative
energy oscillator respectively. At the same time,  from (\ref{z})
recall that our true concern should be with
the $z,p_z$ variables. These are related to  $X,P,x,p$ by the inverse
transformations of (5):
\be
z=\frac{x-(P/\Omega)}{{\sqrt{\Omega^2-\omega ^2}}},~p_z=\frac{
\Omega^2p-\Omega\omega ^2x}{{\sqrt{\Omega^2-\omega ^2}}},$$$$
q=\frac{\Omega X-p}{{\sqrt{\Omega^2-\omega ^2}}},~
p_q=\frac{\omega P-\omega ^2x}{{\sqrt{\Omega^2-\omega ^2}}}.
\label{zpz1}
\ee
Hence it is straightforward to compute $<z>$,
\be
<z>=\frac{1}{{\sqrt{\Omega^2-\omega^2}}}(<x>-\frac{<P>}{\Omega})$$$$=\frac
{1}{{\sqrt{\Omega^2-\omega^2}}}
({\sqrt{\frac{2J}{\Omega}}}sin~\Gamma+{\sqrt{\frac{2j}{\omega}}}
sin~\gamma).
\label{zz}
\ee
Utilizing the operator relation
\be
\dot z=\frac{1}{{\sqrt{\Omega^2-\omega ^2}}}(-p+\Omega X),
\label{dz3}
\ee
one can check that the GCS satisfies the constraint $\lambda (\dot z-q)\sim 0$
in (\ref{xz1}):
\be
<\dot z>=\frac{1}{{\sqrt{\Omega^2-\omega^2}}}
({\sqrt{2\omega j}}~cos\gamma+{\sqrt{2\Omega J}}~cos\Gamma )=<q>.
\label{dz1}
\ee
In the last equality we have used (29).

Next we calculate the dispersion  $(\Delta z)^2$,
\be
<z^2>=\frac{1}{(\Omega^2-\omega
^2)}[\frac{1}{2}(\frac{1}{\omega}+\frac{1}{\Omega})+2({\sqrt{(j/\omega)}}
~sin\gamma+{\sqrt{(J/\Omega)}}~sin\Gamma_1))^2],
\label{z2}
\ee
\be
(\Delta z)^2=<z^2>-<z>^2=\frac{1}{2\Omega\omega (\Omega -\omega)},
\label{dz2}
\ee
as well as the dispersion  $(\Delta p_z)^2$,
\be
<p_z>=-\frac{1}{{\sqrt{\Omega^2-\omega^2}}}(\Omega^2{\sqrt{2\omega j}}
~cos\gamma+\omega ^2{\sqrt{2\Omega J}}~cos\Gamma),
\label{pz}
\ee
\be
<p_z^2>=\frac{1}{(\Omega^2-\omega ^2)}[\frac{1}{2}
\Omega\omega (\omega^3+ \Omega^3) +2(\omega ^2{\sqrt{\Omega J}}
~cos\Gamma
+\Omega^2{\sqrt{\omega j}}~cos\gamma)^2],
\label{pz1}
\ee
\be
(\Delta
p_z)^2=\frac{\Omega\omega (\omega^2+ \Omega^2-\omega \Omega)}{2(\Omega-\omega
)}.
\label{dpz}
\ee
Hence the modified Heisenberg Uncertainty Relation for the physical $z,p_z$
variables is
revealed:
\be
(\Delta z)^2(\Delta
p_z)^2=\frac{(\omega^2+ \Omega^2-\omega \Omega)}{
4 (\Omega-\omega)^2}.
\label{dpz1}
\ee
\vskip .3cm
{\it{Discussions and Conclusion}}: There are several interesting points and peculiarities  to be noticed in the
behavior of the PU Oscillator variable $z$.
Due to the transformations (29) the dimensions of $z,p_z$ are
different from their counterparts $X,P$ or
$x,p$. Furthermore due to the coordinate-momentum mixing in the transformations
(29), the profiles of $<z>,
<p_z>,<z^2>,<p_z^2>$  in (30,33,36) are quite involved with the parameters mixed
up, as compared to the corresponding forms of for a ghost or  normal HO
 $<x>,<p>,<X>,<P>$ in (23-26). However, things get miraculously cleared up once
the
dispersions $(\Delta z)^2,(\Delta p_z)^2$ are computed in (\ref{dz2},\ref{dpz}).
Since the relations
are independent of $\gamma =\omega t, \Gamma =\Omega t$, the dispersions are
time invariant, similar to normal
Harmonic Oscillator. This indicates stability of the GCS.

For $\Omega >>\omega
$, we find,
\be
(\Delta z)^2\sim \frac{1}{\omega \Omega^2}+\frac{1}{\Omega^3}~,~(\Delta
p_z)^2\sim \omega \Omega^2.
\label{dis}\ee
The above immediately provides the leading order correction in the Uncertainty
Relation,
\be
(\Delta z)^2(\Delta p_z)^2\sim \frac{1}{4}(1+\frac{\omega}{\Omega}).
\label{ap}\ee
The energy of the GCS for PU Oscillator will be
\be
E=\frac{\Omega}{2}(1+2J)-\frac{\omega}{2}(1+2j).
\label{en}\ee
Since the parameter $J$ or $j$ can be identified with $\mid z\mid^2$ \cite{gcs1}
it is probably natural to consider $J=j$. In that case the GCS energy is
positive for $\Omega >\omega$ that is
being assumed here.

Finally we comment on the question regarding the stability of the system of two
{\it{interacting}} oscillators
having positive and negative energy levels. Indeed, generically the system will
be unstable but as Smilga \cite{sm,sm2} has
shown that there are certain specific form of interactions for which the system
is stable. However any
interaction term will clearly lead to a non-minimal form of Pais-Uhlenbeck
Oscillator which is not our concern in the present work. However it will be
interesting
to see how effect of interactions  is reflected in the coherent states that we
have constructed here. We expect to report on this in
near future.

To conclude, we have studied the non-degenerate version of the Pais-Uhlenbeck
Oscillator. We suggest that Generalized Coherent States are probably better
suited to deal with the Pais-Uhlenbeck
Oscillator. From previous works \cite{ben,ben2,sm,sm2,mos} it is clear that in spite of
the
presence of negative energy ghost
states the system can be subjected to a consistent quantization program. Our
system lives entirely in positive norm
Hilbert space but we allow presence of negative energy states and hence the
vacuum is unbounded from below. Hence our
main concern is the stability and energy positivity of the coherent states. We
have precisely established that
the coherent states constructed here for Pais-Uhlenbeck
Oscillator can have positive energy under reasonable assumptions on the
parameters and the states have constant
coordinate and momentum dispersions ensuring their stability.

\end{document}